\documentclass[hidelinks,onefignum,onetabnum]{siamart250211}

\usepackage{lipsum}
\usepackage{amsfonts}
\usepackage{graphicx}
\usepackage{epstopdf}
\usepackage{algorithmic}
\usepackage{epsfig}
\ifpdf
  \DeclareGraphicsExtensions{.eps,.pdf,.png,.jpg}
\else
  \DeclareGraphicsExtensions{.eps}
\fi

\newsiamremark{remark}{Remark}
\newsiamremark{hypothesis}{Hypothesis}
\crefname{hypothesis}{Hypothesis}{Hypotheses}
\newsiamthm{claim}{Claim}
\newsiamremark{fact}{Fact}
\crefname{fact}{Fact}{Facts}

\headers{Chemotaxis sensitivity via neural networks}{R. Erban}

\title{Neural networks for learning macroscopic chemotactic sensitivity from microscopic models
}

\author{Radek Erban$\,$\thanks{Mathematical Institute, University of Oxford, Andrew Wiles Building, Radcliffe Observatory Quarter, Woodstock 
Road, Oxford OX2 6GG, United Kingdom (\email{radek.erban@maths.ox.ac.uk}).}}

\usepackage{amsopn}

\begin{document}

\maketitle

\begin{abstract} \noindent
The macroscopic (population-level) dynamics of chemotactic cell movement -- arising from underlying microscopic (individual-based) models -- are often described by parabolic partial differential equations (PDEs) governing the spatio-temporal evolution of cell concentrations. In certain cases, these macroscopic PDEs can be analytically derived from microscopic models, thereby elucidating the dependence of PDE coefficients on the parameters of the underlying individual-based dynamics. However, such analytical derivations are not always feasible, particularly for more complex or nonlinear microscopic models. In these instances, neural networks offer a promising alternative for estimating the coefficients of macroscopic PDEs directly from data generated by microscopic simulations. In this work, three microscopic models of chemotaxis are investigated. The macroscopic chemotaxis sensitivity is estimated using neural networks, thereby bridging the gap between individual-level behaviours and population-level descriptions. The results are compared with macroscopic PDEs, which can be derived for each model in certain parameter regimes.
\end{abstract}



\section{Introduction}

The (classical, Keller-Segel) chemotaxis equation is a partial differential equation (PDE) describing the time evolution of the spatio-temporal concentration of cells $c({\mathbf x},t)$ in response to a spatio-temporal chemical signal $S({\mathbf x},t)$. Assuming that the signal $S({\mathbf x},t)$ is a chemoattractant, the chemotaxis equation can be written as a drift-diffusion equation in domain $\Omega \subset {\mathbb R}^d$, for dimension $d \in \{1,2,3\}$, in the following form~\cite{Horstmann:2003:UPK,Erban:2005:STS}
\begin{equation}
\frac{\partial c}{\partial t}
=
\nabla \cdot
\big( D
\nabla c
-
\, c  \, \chi(S) \, \nabla S
\big),
\label{chemotaxisequation}
\end{equation}
where $c: \Omega \times [0,\infty) \to [0,\infty)$ is the concentraction of cells, $D$ is the diffusion constant and $\chi: [0,\infty) \to [0,\infty)$ is the chemotactic sensitivity. 

In some applications, cells not only detect the chemical $S({\mathbf x},t)$ but also have the capacity to produce it as a signalling molecule or to consume it as a source of nutrition~\cite{Adler:1966:CB,Fisher:1989:QAC}. Then the classical chemotaxis equation is coupled with a PDE describing the dynamics of the chemoattractant~\cite{Hillen:2009:UGP,Bellomo:2015:TMT,Lorenzi:2025:PSC}. An example includes the original Keller-Segel system~\cite{Keller:1971:TBC}, which describes the chemoattractant by a parabolic PDE and uses singular chemotactic sensitivity $\chi(S) = \mathcal{O}(1/S)$ as $S \to 0_+$. Such singularities are necessary to obtain travelling wave solutions in the PDE systems of the Keller-Segel type~\cite{Horstmann:2004:UPK}, which are also sometimes called the Patlak-Keller-Segel PDEs in the literature acknowledging the earlier work of Patlak~\cite{Patlak:1953:RWP}. Both parabolic and elliptic equations have been used for describing the chemoattractant $S({\mathbf x},t)$, resulting in parabolic-parabolic and parabolic-elliptic Keller-Segel PDE systems~\cite{Winkler:2013:FTB,Fujie:2016:GEB2,Arumugam:2021:KSC,Winkler:2022:UGL}. While the chemotactic sensitivity $\chi(S)$ in some theoretical studies is assumed to be constant (independent of $S$), this can result in finite-time blow-up of solutions~\cite{Winkler:2013:FTB}. This can be prevented and the resulting PDE systems can better model biological phenomena by assuming a more realistic functional dependence of the chemotactic sensitivity $\chi(S)$ on the signal $S$~\cite{Ahn:2019:GWP,Fujie:2015:BSP,Fujie:2016:GEB}. For example, common assumptions include that cells are not sensitive to large signal concentrations, and it is required that $\chi(S) \to 0$ as $S \to \infty$ to get globally well-posed systems of PDEs in some models~\cite{Ahn:2019:GWP}.

Chemotaxis can also be modelled using microscopic (individual-based) models. In this paper, we will investigate three such models, denoted as Microscopic Models~I, II and~III, each describing a collection of $N$ individuals (unicellular organisms) and their responses to extracellular chemical signal $S$. Microscopic Model~I is a signal-dependent position-jump process (Brownian dynamics), where the position of each individual is described by a stochastic differential equation (SDE). Microscopic Model~II is a signal-dependent velocity-jump process, while the most complex Microscopic Model~III also includes internal (intracellular) dynamics describing signal processing by each individual. We investigate the problem of inferring macroscopic PDEs, including the functional form of the chemotactic sensitivity $\chi(S)$, for each microscopic model. Such PDEs can be analytically derived under some assumptions for simplified individual-based models~\cite{Othmer:1988:MDB,Erban:2004:ICB,Xue:2015:MEB,Franz:2016:HIV,Mayo:2025:BCC}. We will show that the macroscopic chemotaxis sensitivity can also be estimated using feedforward neural networks~\cite{Goodfellow:2016:DL}. 

The paper is organised as follows. In Section~\ref{sec2}, we introduce the chemotaxis equation, its numerical discretisation and our feedforward neural network architecture. Microscopic Models~I, II and~III are presented, and their macroscopic description inferred in Sections~\ref{sec3}, \ref{sec4} and~\ref{sec5}, respectively. For each model, we will discuss parameter regimes where their macroscopic behaviour is well approximated by the parabolic chemotaxis equation~(\ref{chemotaxisequation}). Considering velocity-jump processes, their macroscopic description can also be obtained in the form of the hyperbolic chemotaxis equation~\cite{Hillen:2002:HMC,Filbet:2005:DHM,Erban:2006:GER}. We will show in Sections~\ref{sec4} and~\ref{sec5} that the loss function based on the hyperbolic chemotaxis equation is more suitable for estimating the chemotactic sensitivity from relatively short simulations of individual-based models. We conclude with a discussion of our results in Section~\ref{sec6}.

\section{Chemotaxis equation and neural networks}

\label{sec2}

\begin{figure}
\rule{0pt}{1pt}
\vskip 1mm
\epsfig{file=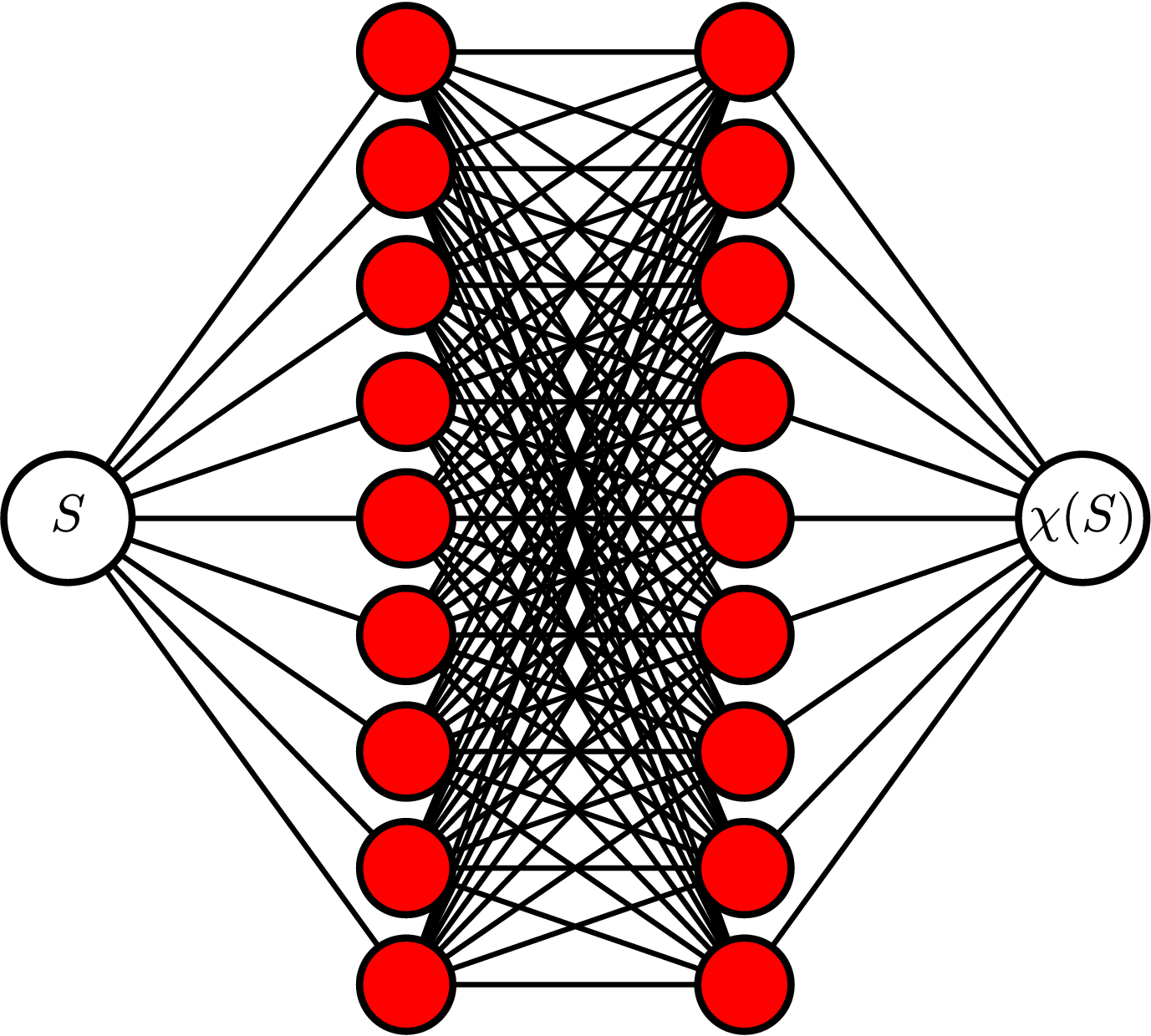,height=5.3cm}
\hskip 6mm 
\raise -1mm
\hbox{\epsfig{file=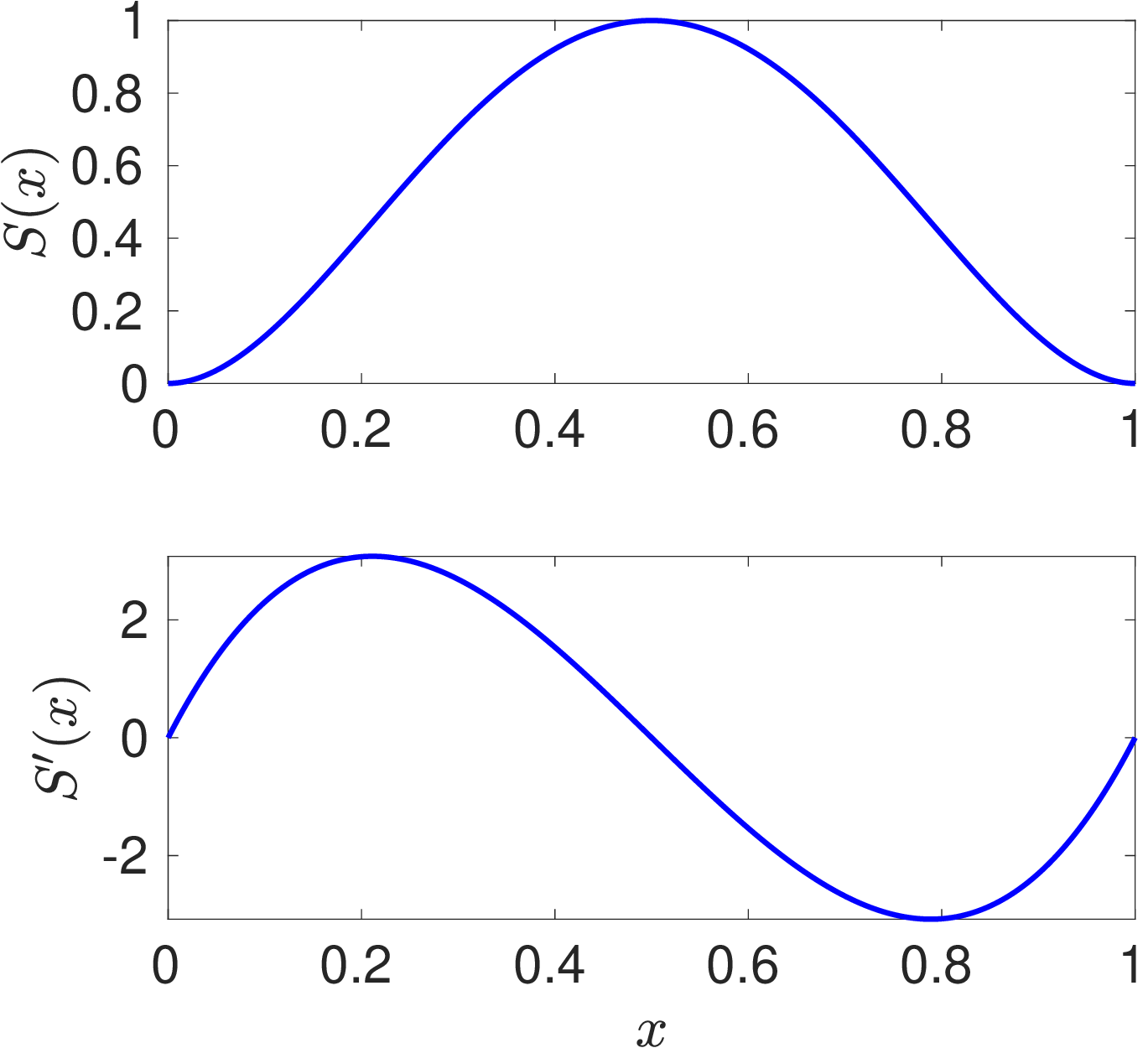,height=5.8cm}}
\vskip -6cm
(a) \hskip 13mm hidden layers \hskip 23mm (b)
\vskip 5.6cm
\caption{{\it {\rm (a)} Schematic of the feedforward neural network architecture with two hidden layers. \hfill\break {\rm (b)} The chemotactic signal $S(x)$ $($top panel$\,)$ and its derivative $S^\prime(x)$ $($bottom panel$\,)$ given by equations~$(\ref{signalS})$ and~$(\ref{dersignalS})$.
}}
\label{fig1}
\end{figure}

In this paper, we will restrict our consideration to spatially one-dimensional chemotaxis models, with the macroscopic concentration of cells denoted by $c(x,t)$, where $x \in [0,L]$ for $L > 0.$ In particular, domain $\Omega$ is the one-dimensional interval, $\Omega = [0,L]$, and the macroscopic chemotaxis equation (\ref{chemotaxisequation}) can be rewritten in the form
\begin{equation}
\frac{\partial c}{\partial t}
=
D \, \frac{\partial^2 c}{\partial x^2}
-
\frac{\partial}{\partial x}
\left(
c \, \chi(S) \, \frac{\partial S}{\partial x}
\right).
\label{1Dchemeqn}
\end{equation}
We will assume that the signal profile varies in space but remains independent of time, i.e. $S \equiv S(x)$. Our objective is to infer the appropriate form of chemotactic sensitivity $\chi(S)$, or, more generally, the entire macroscopic chemotaxis equation, based on the estimation of cell concentration $c(x,t)$ by microscopic models in Sections~\ref{sec3}, \ref{sec4} and~\ref{sec5}. We discretise interval $[0,L]$ using $n \in {\mathbb N}$ meshpoints at locations
\begin{equation}
\mathbf{x}_{i}
=
(i-1/2) \, \Delta x,
\qquad 
i=1,2,\dots,n,
\qquad
\mbox{where} \quad
\Delta x =  \frac{L}{n}\,,
\label{meshpoints}
\end{equation}
and denote
\begin{equation}
c_{i}(t)
=
c(x_{i},t),
\qquad
S_{i}
=
S(x_{i}).
\label{cisikdef}
\end{equation}
Then concentrations $c_i(t)$, $i=1,2,\dots,n$, can be estimated in our simulations of individual-based models in Sections~\ref{sec3}, \ref{sec4} and~\ref{sec5} by calculating the number of individuals in interval $[(i-1)\Delta x, i\Delta x]$ at time $t$. The notation~(\ref{cisikdef}) can also be used to describe a numerical method for solving the macroscopic chemotaxis equation~(\ref{1Dchemeqn}). Using finite differences to discretise the right-hand side of equation~(\ref{1Dchemeqn}), we obtain
\begin{equation}
\frac{\mbox{d} c_{i}}{\mbox{d} t}
=
{\mathcal A}_{i} (c,S)
\label{chemodiscr}
\end{equation}
where ${\mathcal A}_{i} (c,S)$ is defined by
\begin{eqnarray}
{\mathcal A}_{i} (c,S)
&=&
\frac{D}{\Delta x^2}
\,
\big( c_{i+1} - 2 c_{i} + c_{i-1} \big)
\, + \,
\chi\!\left(\!\frac{S_{i+1} + S_{i}}{2} \!\right)
\frac{\big(c_{i+1} + c_{i}\big) \big(S_{i} - S_{i+1}\big)}{2 \, \Delta x^2} 
\nonumber
\\
&& +
\chi\!\left(\!\frac{S_{i-1} + S_{i}}{2} \!\right)
\frac{\big(c_{i-1} + c_{i}\big) \big(S_{i} - S_{i-1}\big)}{2 \, \Delta x^2}\,,
\qquad
\label{defai}
\end{eqnarray}
for internal mesh points $i=2,3,\dots,n-1$. To extend the definition of ${\mathcal A}_{i} (c,S)$ to boundary mesh points at $i=1$ and $i=n$, we need to specify the boundary conditions. In our simulations of individual-based models in Sections~\ref{sec3}, \ref{sec4} and~\ref{sec5}, we will use reflective (zero-flux) boundary conditions, which preserve the number of simulated individuals and correspond, at the macroscopic level, to boundary terms
\begin{eqnarray}
{\mathcal A}_{1} (c,S)
&=&
\frac{D}{\Delta x^2}
\,
\big( c_{2} - c_{1} \big)
\, + \,
\chi\!\left(\!\frac{S_{2} + S_{1}}{2} \!\right)
\frac{\big(c_{2} + c_{1}\big) \big(S_{1} - S_{2}\big)}{2 \, \Delta x^2}\,, 
\label{defa1}
\\
{\mathcal A}_{n} (c,S)
&=&
\frac{D}{\Delta x^2}
\,
\big( c_{n-1} - c_{n} \big)
\, + \,
\chi\!\left(\!\frac{S_{n-1} + S_{n}}{2} \!\right)
\frac{\big(c_{n-1} + c_{n}\big) \big(S_{n} - S_{n-1}\big)}{2 \, \Delta x^2}\,.
\label{defan}
\end{eqnarray}
Denoting the integral of $\chi$ by $\psi$, we have $\psi^\prime = \chi$ and equation~(\ref{1Dchemeqn}) can be rewritten
as
$$
\frac{\partial c}{\partial t}
=
D \, \frac{\partial^2 c}{\partial x^2}
-
\frac{\partial}{\partial x}
\left(
c \, \frac{\partial \mbox{\hskip 0.2mm}\psi(S)}{\partial x}
\right).
$$
The steady state solution of the macroscopic chemotaxis equation~(\ref{1Dchemeqn}) can then be written as
\begin{equation}
c_s(x)
\,=\,
\lim_{t \to \infty}
c(x,t)
\,=\,
A \exp\!\left[\frac{\psi(S)}{D}\right]
\label{ststsol}
\end{equation}
where $A>0$ is a constant. In particular, if the macroscopic equation is described by the chemotaxis equation~(\ref{1Dchemeqn}), then the knowledge of the equilibrium (steady state) concentration profile $c_s(x)$ and the signal profile $S(x)$ could be used to infer the chemotactic sensitity $\chi(S)$ by plotting the signal $S$ against
$$
\chi(S) 
\, = \,
\frac{D \, c_s^\prime(x)}{c_s(x) \, S^\prime(x)} \, ,
$$
where primes denote derivatives. However, such an approach would require running long-time simulations of macroscopic models until the equilibrium can be sampled. We will instead adopt an approach where we will evolve microscopic models over a specific time interval. We will consider transient time-dependent data obtained by microscopic models in Sections~\ref{sec3}, \ref{sec4} and~\ref{sec5}, where we calculate the concentration profile $c(x,t+\tau)$ at time $t+\tau$ by evolving the microscopic model over the time interval $[t,t+\tau]$, where $\tau>0$. This will be used to estimate the time derivative in equation~(\ref{chemodiscr}) as
$$
 \frac{\mbox{d} c_{i}}{\mbox{d} t}
=
\frac{c_i(t+\tau) - c_i(t)}{\tau} \, .
$$
Chemotaxis sensitivity $\chi(S)$ will be approximated by feedforward neural network architecture~\cite{Goodfellow:2016:DL}, which is schematically shown in Figure~\ref{fig1}(a). The architecture is structured to accept a single input, the value of $S$, which is managed by the first fully connected layer. This layer is designed to transform the input size of one into an output size of $n_1$, effectively mapping the single-dimensional input into a higher-dimensional space. This is achieved by applying an affine transformation ${\mathbf w}_1 S + \mathbf{b}_1$, where ${\mathbf w_1} \in {\mathbb R}^{n_1}$ and ${\mathbf b_1} \in {\mathbb R}^{n_1}$ are the weight vector and the bias vector, respectively. Following this transformation, the non-linearity is introduced through the application of a ReLU activation function defined by
\begin{equation}
\mbox{ReLU}(z) \, = \, \max \{0,z\} \, = \, \left\{ \begin{matrix} z \quad & \mbox{for} \quad z > 0 \,; \\ 0 \quad & \mbox{for} \quad z \le 0\,. \end{matrix} \right. 
\label{defReLU}
\end{equation} 
This non-linear activation enables the neural network to capture more complex relationships within the data. 

Our neural network architecture in Figure~\ref{fig1}(a) continues with a second hidden layer, which mirrors the structure of the first hidden layer. It takes the output from the previous layer, ${\mathbf z}_1 \in {\mathbb R}^{n_1}$, and again applies an affine transformation, which can be written as $W_2 \, {\mathbf z}_1 + {\mathbf b_2}$, where $W_2 \in {\mathbb R}^{n_2 \times n_1}$ is the weight matrix, ${\mathbf b_2} \in {\mathbb R}^{n_2}$ is the bias vector, and $n_2$ is the size of the second hidden layer. As with the first layer, the ReLU activation function~(\ref{defReLU}) is applied to the output of the second layer, ensuring that the model retains the ability to learn intricate, non-linear relationships. Finally, the architecture in Figure~\ref{fig1}(a) culminates in the output layer, which is responsible for condensing the information down to a single output value. This layer takes the $n_2$-dimensional input from the second hidden layer, ${\mathbf z}_2 \in {\mathbb R}^{n_2}$, and applies an affine transformation that results in a single output size of one, namely we use ${\mathbf w}_3 \cdot {\mathbf z}_2 + b_3$, where ${\mathbf w_3} \in {\mathbb R}^{n_2}$ and $b_3 \in {\mathbb R}$ are the weight vector and the bias, respectively. To train the network, we use equation~(\ref{chemodiscr}) to define the loss function
\begin{equation}
{\mathcal L}(S)
\,=\,
\frac{1}{n}
\sum_{i=1}^n
\left(
\frac{c_i(t+\tau) - c_i(t)}{\tau}
\,-\,
{\mathcal A}_{i} (c,S)
\!\right)^{\!2} \, ,
\label{lossfunction}
\end{equation}
and we use the Adam optimizer to minimize prediction errors based on the training dataset for each microscopic model. We implement our feedforward neural network architecture using PyTorch.

In our illustrative computational examples in Sections~\ref{sec3}, \ref{sec4} and~\ref{sec5}, we will assume that all parameters have already been non-dimensionalized. We use the domain size $L=1$ and the chemotactic signal profile $S : \Omega \to [0,1]$ given by
\begin{equation}
S(x) \, = \, 2^4 \, x^2 (x-1)^2\,.
\label{signalS}
\end{equation}
This signal profile is plotted in the top panel of Figure~\ref{fig1}(b). Differentiating~(\ref{signalS}), we obtain
\begin{equation}
S^\prime(x) \, = \, 2^{5} x (2 x - 1 ) (x-1)
\qquad \mbox{and} \qquad 
S^{\prime\prime}(x) \, = \, 2^{5} (6 x^2 - 6x + 1)\,.
\label{dersignalS}
\end{equation}
In particular, considering the domain $\Omega=[0,1]$, the signal $S(x)$ has minima at $x=0$ and $x=1$ equal to $S(0)=S(1)=0$ and its maximum at $x=1/2$ equal to $S(1/2)=1$. The derivative $S^\prime(x)$ is plotted in the bottom panel of Figure~\ref{fig1}(b). It satisfies 
$$
|S^\prime(x)| \, \le \, \frac{16}{3 \sqrt{3}} \, \doteq \, 3.0792 \dots
\qquad
\mbox{for}
\quad
x \in \Omega=[0,1]\,.
$$

\section{Microscopic Model I (Brownian dynamics)}

\label{sec3}

\begin{figure}
\rule{0pt}{1pt}
\vskip 1mm
\epsfig{file=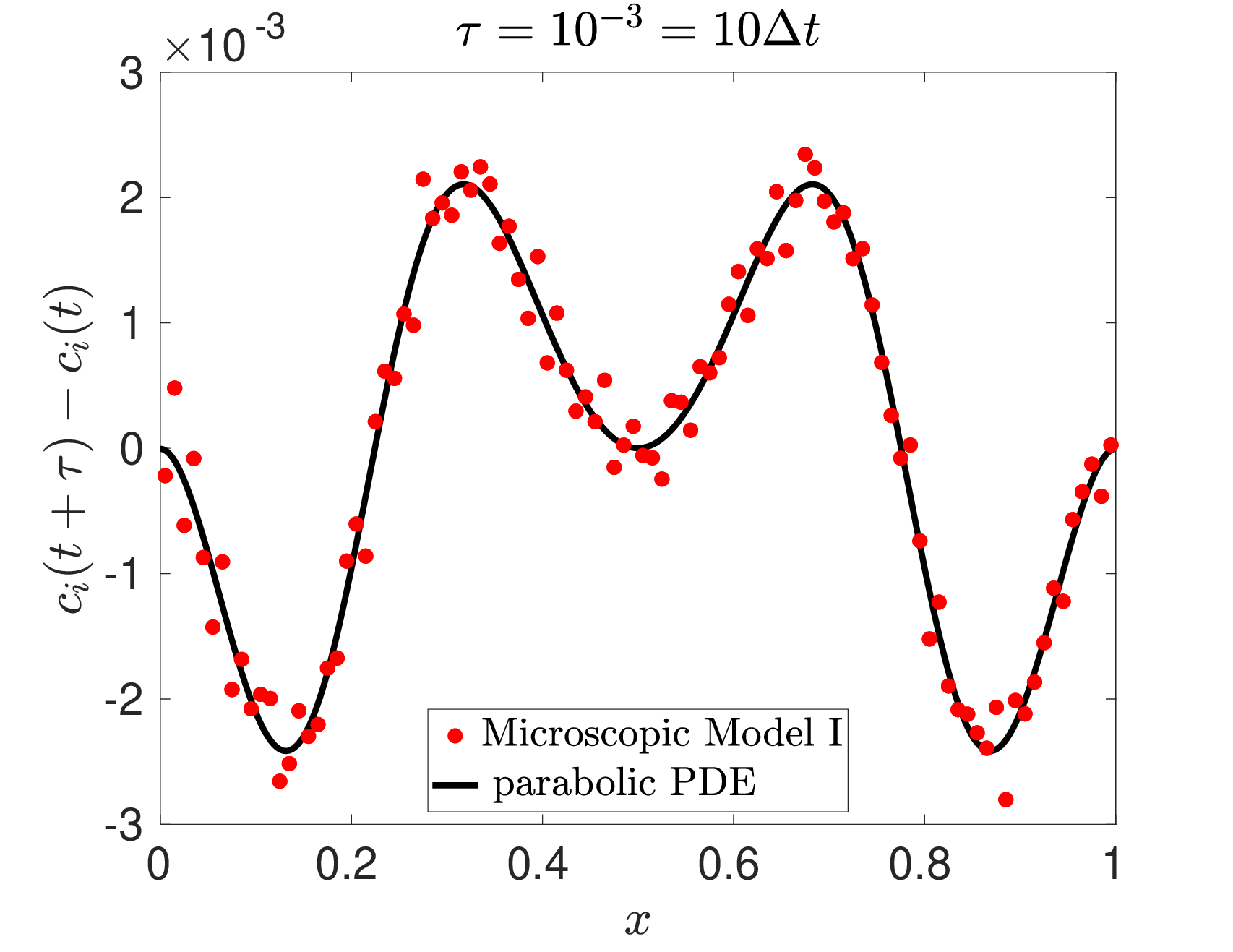,height=5.1cm}
\hskip -2.8mm
\epsfig{file=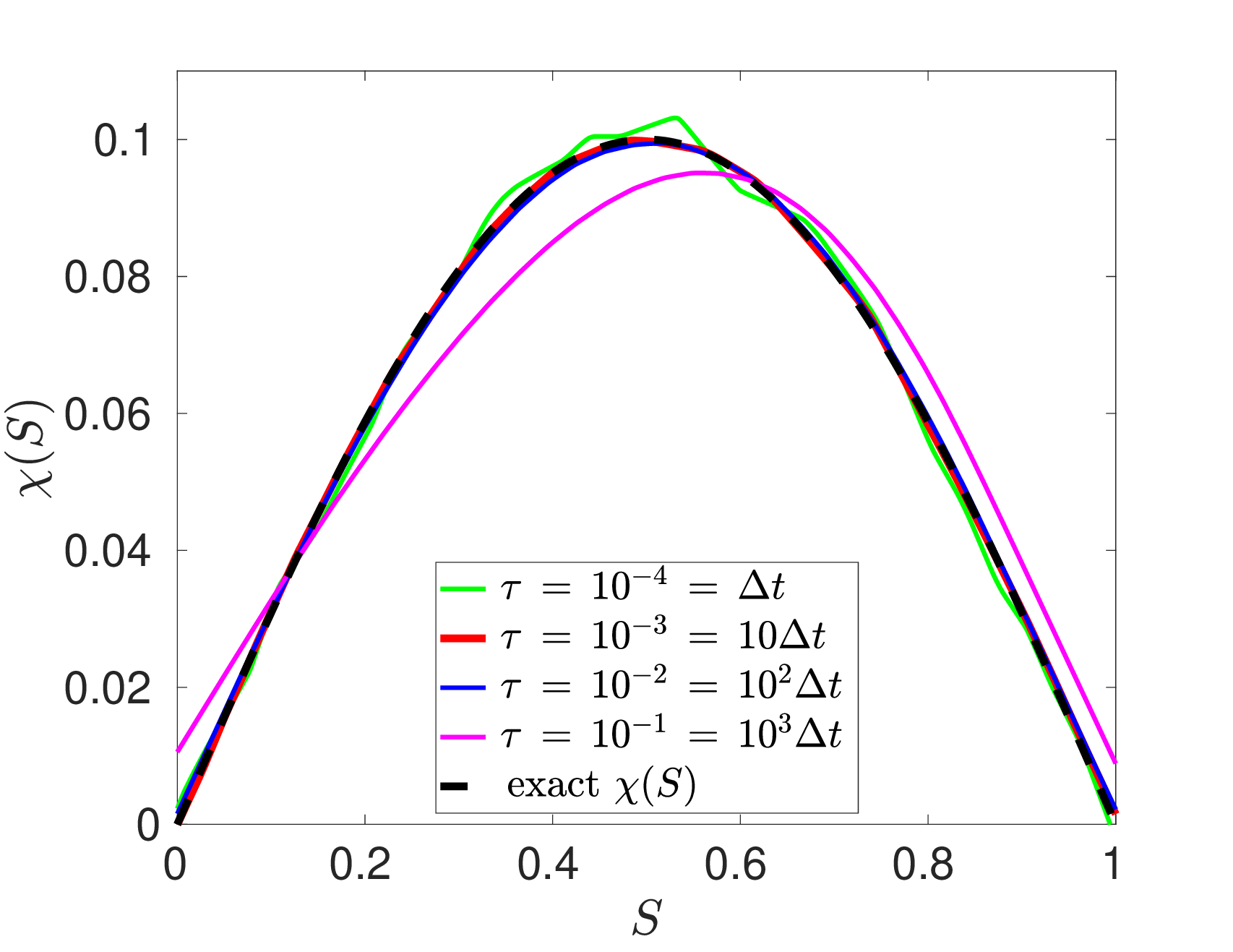,height=5.1cm}
\vskip -5.3cm
(a) \hskip 59mm (b)
\vskip 4.7cm
\caption{{\it {\rm (a)} The change in the concentration $c_i(t+\tau)-c_i(t)$ over the time interval of length $\tau$, starting from the uniform concentration profile $c_i(t) \equiv 1$, for $i=1,2,\dots,n$, calculated by Microscopic model $\,${\rm I} $($red circles$\,)$. We use chemotactic sensitivity~$(\ref{givenchiS})$, signal profile~$(\ref{signalS})$, $N=10^9$ cells in the microscopic model, time interval $\tau = 10^{-3} = 10 \Delta t$, $D=10^{-2}$ and $n=100$. The solution of the parabolic chemotaxis equation~$(\ref{1Dchemeqn})$ is plotted as the black line.
 \hfill\break {\rm (b)} The chemotactic sensitivity estimated by the feedforward neural network for different values of time window $\tau$. The exact chemotactic sensitivity~$(\ref{givenchiS})$ is plotted as the black dashed line.
}}
\label{fig2}
\end{figure}

We model a system of $N$ individuals (unicellular organisms) in domain $\Omega = [0,L]$, where $L > 0$. The state of the system at time $t$ is described by the $N$-dimensional vector of positions
\begin{equation}
\mathbf{X}(t)
=
\big[X_1(t), X_2(t), \dots, X_N(t)\big]
\in \Omega^{N},
\label{statevector}
\end{equation}
where $X_i(t)$ denotes the position of the $i$-th cell at time $t$ for $i=1,2,\dots,N$. The time evolution of $X_i(t)$ is given by the following It\^{o} SDE
\begin{equation}
\mbox{d} X_i(t)
= 
\chi(S) \, \frac{\partial S}{\partial x}
\,
\mbox{d}t
\, + \, 
\sqrt{2D} \,
\mbox{d} W_i,
\qquad \mbox{for} \quad
i=1,2,\dots,N.
\label{BDSDEform}
\end{equation}
Then the chemotaxis equation (\ref{1Dchemeqn}) is exactly equal to the Fokker-Planck equation describing the time evolution of the spatio-temporal probability density of each individual~\cite{Erban:2020:SMR}. In particular, if we simulate the microscopic model and calculate the number of individuals in interval $[(i-1)\Delta x, i\Delta x]$ at time $t$ to estimate the density $c_i(t) = c(x_i,t)$ given by~(\ref{cisikdef}), then we are effectively calculating (noisy) solutions of the chemotaxis equation~(\ref{1Dchemeqn}). The error (noise) reduces to zero in the large particle limit $N \to \infty$ provided that the equation~(\ref{BDSDEform}) is solved exactly. In practice, we simulate a finite number of individuals, $N$, and we discretise the SDE~(\ref{BDSDEform}) using a finite time step $\Delta t$ to obtain
\begin{equation}
X_i(t+\Delta t)
\,=\, 
X_i(t)
+
\chi(S) \, \frac{\partial S}{\partial x}
\,
\Delta t
\, + \, 
\sqrt{2D \, \Delta t} \,
\xi_i,
\qquad \mbox{for} \quad
i=1,2,\dots,N,
\label{BDSDEformdiscr}
\end{equation}
where $\xi_i$ is normally distributed random number with the zero mean and unit variance. In our simulations, we consider $N=10^9$ individuals, $D=10^{-2}$, $\Delta t = 10^{-4}$ and the signal profile given by~(\ref{signalS}) together with the chemotactic sensitivity 
\begin{equation}
\chi(S) = \frac{\sin(\pi S)}{10}\,.
\label{givenchiS}
\end{equation}
Since our signal profile~(\ref{signalS}) satisfies $0 \le S(x) \le 1$ in interval $\Omega = [0,1]$, we have $0 \le \chi(S) \le 1/10$ with the highest chemotactic sensitivity achieved for intermediate signal values at $S = 1/2$. The steady state solution~(\ref{ststsol}) is given by
\begin{equation}
c_s(x)
\,=\,
A \exp\!\left[\frac{-\cos(\pi S(x))}{10 \pi D}\right]
\,=\,
A \exp\!\left[\frac{-\cos(2^4 \, \pi \, x^2 (x-1)^2)}{10 \pi D}\right],
\label{ststsol2}
\end{equation}
where $A>0$ is a constant. The steady state solution~(\ref{ststsol2}) has its maximum at $x=1/2$ and concentration $c(x,t)$ will approach $c_s(x)$, provided that we observe the system for sufficiently long time. Considering transient dynamics, concentration $c(x,t)$ will, in general, not have its maximum at $x=1/2$. This is illustrated in Figure~\ref{fig2}(a), where we consider that cells are uniformly distributed at time $t$, i.e. $c_i(t) \equiv 1$, and we calculate their time evolution by~(\ref{BDSDEformdiscr}) over the time interval $[t,t+\tau]$, where $\tau = 10^{-3} = 10 \Delta t$. In Figure~\ref{fig2}(a), we plot the estimated change in concentration $c_i(t+\tau)-c_i(t) \equiv c_i(t+\tau)-1$ and compare it with the solution of the parabolic chemotaxis equation (\ref{1Dchemeqn}). We observe that there are initially no significant changes in concentration at points $x=0$, $x=1/2$ and $x=1$, which correspond to minima (at $x=0$ and $x=1$) and maximum (at $x=1/2$) of the signal $S(x)$.  

Next, we use data on positions of cells calculated by relatively short simulations of Microscopic Model~I to estimate the chemotactic sensitivity $\chi(S)$. We use the feedforward neural network architecture schematically shown in Figure~\ref{fig1}(a) with $n_1=n_2=50$. Our training data are calculated as in Figure~\ref{fig2}(a), where we consider that cells are uniformly distributed at time $t$, i.e. $c_i(t) \equiv 1$, and we calculate their time evolution by applying equation~(\ref{BDSDEformdiscr}) over the time interval $[t,t+\tau]$ to estimate $c_i(t+\tau)$, where $\tau>0$. We use 
$$
\tau \in \big\{ \Delta t, \, 10 \Delta t, \, 10^2 \Delta t, \, 10^3 \Delta t \big\}
$$ 
and train 100 neural networks (using $10^3$ epochs) for each pair $c_i(t)$ and $c_i(t+\tau)$. The results are presented in Figure~\ref{fig2}(b), where we average the calculated chemotactic sensitivity $\chi(S)$ over 100 realizations of the training process for each value of $\tau$. 

Comparing the neural network estimates of $\chi(S)$ in Figure~\ref{fig2}(b) with the exact result~(\ref{givenchiS}), we observe that the best results are obtained when using intermediate values $\tau=10 \Delta t$ and $\tau=10^2 \Delta t$. On one hand, if $\tau$ is small (i.e. for $\tau = \Delta t$ in Figure~\ref{fig2}(b)), then the results are more noisy (even after averaging over 100 realizations). This is caused by more noisy training data for small values of $\tau$. On the other hand, if $\tau$ is larger (i.e. for $\tau = 10^3 \Delta t$ in Figure~\ref{fig2}(b)), then we observe a systematic bias in Figure~\ref{fig2}(b), caused by inaccurate estimation of the time derivative in the loss function~(\ref{lossfunction}). 

\section{Microscopic Model~II (velocity-jump process)}

\label{sec4}

To introduce more microscopic detail into the modelling of chemotaxis, we will describe individual cells not only by their positions~(\ref{statevector}), but also by their velocities. Then we can view chemotaxis at the individual-based level as a velocity-jump process~\cite{Othmer:1988:MDB,Hillen:2000:DLT}. Velocity-jump processes have been used for modelling the chemotaxis of flagellated bacteria, such as {\it E. coli}, which alternates between two modes of behaviour: a more or less straight motion with constant speed called ``run" and a highly erratic motion called ``tumble", which produces little translation but reorients the cell~\cite{Erban:2004:ICB}. Since the tumble time is shorter than the average run time, a tumble can be viewed as an instantaneous change in velocity. In this section, we again restrict to one spatial dimension and we model a system of $N$ unicellular organisms in domain $\Omega = [0,L]$, where the state of the system at time $t$ is described by two $N$-dimensional vectors: the vector of positions~(\ref{statevector}) and the vector of velocities denoted by
\begin{equation}
\mathbf{V}(t)
=
\big[V_1(t), V_2(t), \dots, V_N(t)\big]
\in {\mathbb R}^{N},
\label{statevectorvelocity}
\end{equation}
where $V_i(t)$ is the velocity of the $i$-th cell at time $t$ for $i=1,2,\dots,N$. We assume that each cell moves along the $x$-axis at a constant speed $\beta>0$, i.e. its velocity $V_i(t)$ only takes one of two possible values, $V_i(t) = \pm \beta$. Each cell reverses its direction at random instants of time according to the Poisson process with turning frequency
\begin{equation}
\mbox{ReLU} \!\left( \lambda_0 \, \pm \, \frac{\lambda_0 \, \chi(S)}{\beta} \, \frac{\partial S}{\partial x} \right),
\label{turnfreq}
\end{equation}
where $\lambda_0 > 0$, the function $\mbox{ReLU}$ is defined by~(\ref{defReLU}) and the sign $\pm$ depends on the direction of the cell movement: a plus sign is for the individuals moving to the left and a minus sign is for the individuals moving to the right. This choice of turning frequency ensures that the underlying velocity-jump process is biased in a way that the cell is less likely to change direction when moving in a favourable direction, i.e.  in the direction in which the signal function $S$ is increasing. Taking into account that the velocity of each individual, $V_i(t)$, takes only two possible values $V_i(t) = \pm \beta$, the turning frequency~(\ref{turnfreq}) can also be equivalently 
rewritten as
\begin{equation}
\mbox{ReLU} \!\left( \lambda_0 \left( 1 \, - \, \frac{\chi(S(X_i(t)))}{V_i(t)} \, \frac{\partial S}{\partial x} (X_i(t)) \right) \right),
\label{turnfreq2}
\end{equation}
where $\pm$ sign has been adsorbed into the sign of the velocity $V_i(t)$ and we have also explicitly indicated that the signal and its derivative are evaluated at the current position of the individual, $X_i(t)$. The formula~(\ref{turnfreq}) includes the function $\mbox{ReLU}$ to ensure that the turning frequency is nonnegative. However, if the signal $S(x)$ satisfies
\begin{equation}
\left| \frac{\chi(S)}{\beta} \, \frac{\partial S}{\partial x} \right| \, \le \, 1 \,,
\label{smallsignal}
\end{equation}
then we can drop the function $\mbox{ReLU}$ from the turning frequency~(\ref{turnfreq}), because its argument is always nonnegative. Moreover, let $c^+(x,t)$ be the concentration of cells that are at $(x,t)$ and are moving to the right, and let $c^-(x,t)$ be the concentration of cells that are at $(x,t)$ and are moving to the left. Then $c^{\pm} (x,t)$ satisfy the equations
\begin{eqnarray}
\frac{\partial c^+}{\partial t}
+
s
\frac{\partial c^+}{\partial x}
&=&
- \left( \lambda_0 - \frac{\lambda_0 \, \chi(S)}{\beta} \, \frac{\partial S}{\partial x} \right) c^+ 
+ \left( \lambda_0 + \frac{\lambda_0 \, \chi(S)}{\beta} \, \frac{\partial S}{\partial x} \right) c^-,
\label{pplussig}\\
\frac{\partial c^-}{\partial t}
-
s
\frac{\partial c^-}{\partial x}
&=&
\left( \lambda_0 - \frac{\lambda_0 \, \chi(S)}{\beta} \, \frac{\partial S}{\partial x} \right) c^+ 
- \left( \lambda_0 + \frac{\lambda_0 \, \chi(S)}{\beta} \, \frac{\partial S}{\partial x} \right) c^-.
\label{pminussig}
\end{eqnarray}
The concentration of cells at $(x,t)$ is given by the sum $c(x,t) = c^+(x,t) + c^-(x,t).$ 
Adding and subtracting the equations~(\ref{pplussig})--(\ref{pminussig}), we get
\begin{eqnarray}
\frac{\partial c}{\partial t}
\,+\,
\beta\,
\frac{\partial q}{\partial x}
\,&=&\,
0\,,
\label{aux1}\\
\frac{\partial q}{\partial t}
\,+\,
\beta\,
\frac{\partial c}{\partial x}
\,&=&\,
- 2 \, \lambda_0 \, q 
\,+\,
\frac{2 \, \lambda_0 \, \chi(S)}{\beta} \, \frac{\partial S}{\partial x} \, c\,,
\label{aux2}
\end{eqnarray}
where $q(x,t) = c^+(x,t) - c^-(x,t).$ Differentiating equation~(\ref{aux1}) with respect to $t$ and equation~(\ref{aux2})
with respect to $x$, we can eliminate $q$ to deduce that $c(x,t)$ satisfies the second order PDE
\begin{equation}
\frac{1}{2 \lambda_0}
\frac{\partial^2 c}{\partial t^2}
\, + \,
\frac{\partial c}{\partial t}
\, = \, 
D \, \frac{\partial^2 c}{\partial x^2}
-
\frac{\partial}{\partial x}
\left(
c \, \chi(S) \, \frac{\partial S}{\partial x}
\right),
\label{csecorder}
\end{equation}
where we have denoted
\begin{equation}
D 
\, 
= 
\,
\frac{\beta^2}{2 \lambda_0}
\,.
\label{Dform}
\end{equation}
The macroscopic equation~(\ref{csecorder}) is a hyperbolic version of the classical 
chemotaxis equation~(\ref{1Dchemeqn}) containing an additional term with the second 
time derivative. In particular, if we simulate Microscopic Model~II and calculate the 
number of individuals in interval $[(i-1)\Delta x, i\Delta x]$ at time $t$ to estimate 
the density $c_i(t) = c(x_i,t)$ given by~(\ref{cisikdef}), then we are effectively calculating 
(noisy) solutions of the hyperbolic chemotaxis equation~(\ref{csecorder}). The error (noise) 
reduces to zero in the large particle limit $N \to \infty$ provided that the Poisson
process with the turning frequency~(\ref{turnfreq}) is implemented exactly and the
signal $S$ satisfies~(\ref{smallsignal}). Otherwise, we will only obtain an approximation
of a solution to the equation~(\ref{csecorder}). Since the hyperbolic PDE~(\ref{csecorder}) reduces
for sufficiently large times to the parabolic chemotaxis 
equation, we can also use Microscopic Model~II to estimate the
solutions of the classical chemotaxis equation~(\ref{1Dchemeqn}), or its steady state behaviour~\cite{Hillen:2000:HMC}.

To solve equation~(\ref{csecorder}) numerically, we discretise interval $[0,L]$ using $n \in {\mathbb N}$ 
meshpoints at locations~(\ref{meshpoints}) and use notation~(\ref{cisikdef}) as we have done when solving 
the parabolic chemotaxis equation~(\ref{1Dchemeqn}) in Section~\ref{sec2}. Using finite differences 
to discretise the right-hand side of equation~(\ref{csecorder}), we obtain
\begin{equation}
\frac{1}{2 \lambda_0}
\frac{\mbox{d}^2 c_{i}}{\mbox{d} t^2}
\, + \,
\frac{\mbox{d} c_{i}}{\mbox{d} t}
\, = \,
{\mathcal A}_{i} (c,S)\,,
\label{hyperchemodiscr}
\end{equation}
where ${\mathcal A}_{i} (c,S)$, for $i=1,2,\dots,n$, is defined by (\ref{defai})--(\ref{defan}). The system of
$n$ second-order ordinary differential equations (ODEs)~(\ref{hyperchemodiscr}) can be rewritten as a system of $2n$ first-order
ODEs
\begin{eqnarray}
\frac{\mbox{d} c_{i}}{\mbox{d} t}
\, &=& \,
z_i \label{hyperchemodiscrci}
\\
\frac{\mbox{d} z_{i}}{\mbox{d} t}
\, &=& \,
2 \lambda_0 
\, \big(
{\mathcal A}_{i} (c,S)
\,-\,
z_{i}
\big)
\label{hyperchemodiscrzi}
\end{eqnarray}
where the auxiliary variable $z_i$ is the time derivative of $c_i$. Alternatively, we could also use other sets of two variables to rewrite the second-order PDE~(\ref{csecorder}) as a system of two first-order PDEs, as it is shown with variables $c^+$ and $c^-$ in equations~(\ref{pplussig})--(\ref{pminussig}) and variables $c$ and $q$ in equations~(\ref{aux1})--(\ref{aux2}).

To simulate Microscopic Model~II, we update the system over time steps of length~$\Delta t$. Multiplying the turning frequency~(\ref{turnfreq2}) by $\Delta t$ and assuming that $\Delta t$ is sufficiently small, the probability that the $i$-th cell changes its direction during one time step is given by
\begin{equation}
p_i(t)
=
\mbox{ReLU} \!\left( \lambda_0 \left( 1 \, - \, \frac{\chi(S(X_i(t)))}{V_i(t)} \, \frac{\partial S}{\partial x} (X_i(t)) \right) \right) \Delta t \, .
\label{probturning}
\end{equation}
In our simulations, we generate $N$ random numbers $r_i$, $i=1,2,\dots,N$, uniformly distributed in interval $[0,1]$, and the $i$-th cell changes its direction of movement at time $t$ if $r_i < p_i(t)$. Then we update the position $X_i(t)$ of each cell by
\begin{equation}
X_i(t+\Delta t) = X_i(t) + V_i(t) \,\Delta t\,,
\label{positionupdate}
\end{equation}
where $V_i(t)$ is assumed to be constant during the time interval $[t,t+\Delta t$], i.e. we have $V_i(t) = \pm \beta$ with the sign depending on the direction of movement. 

\begin{figure}
\rule{0pt}{1pt}
\vskip 1mm
\epsfig{file=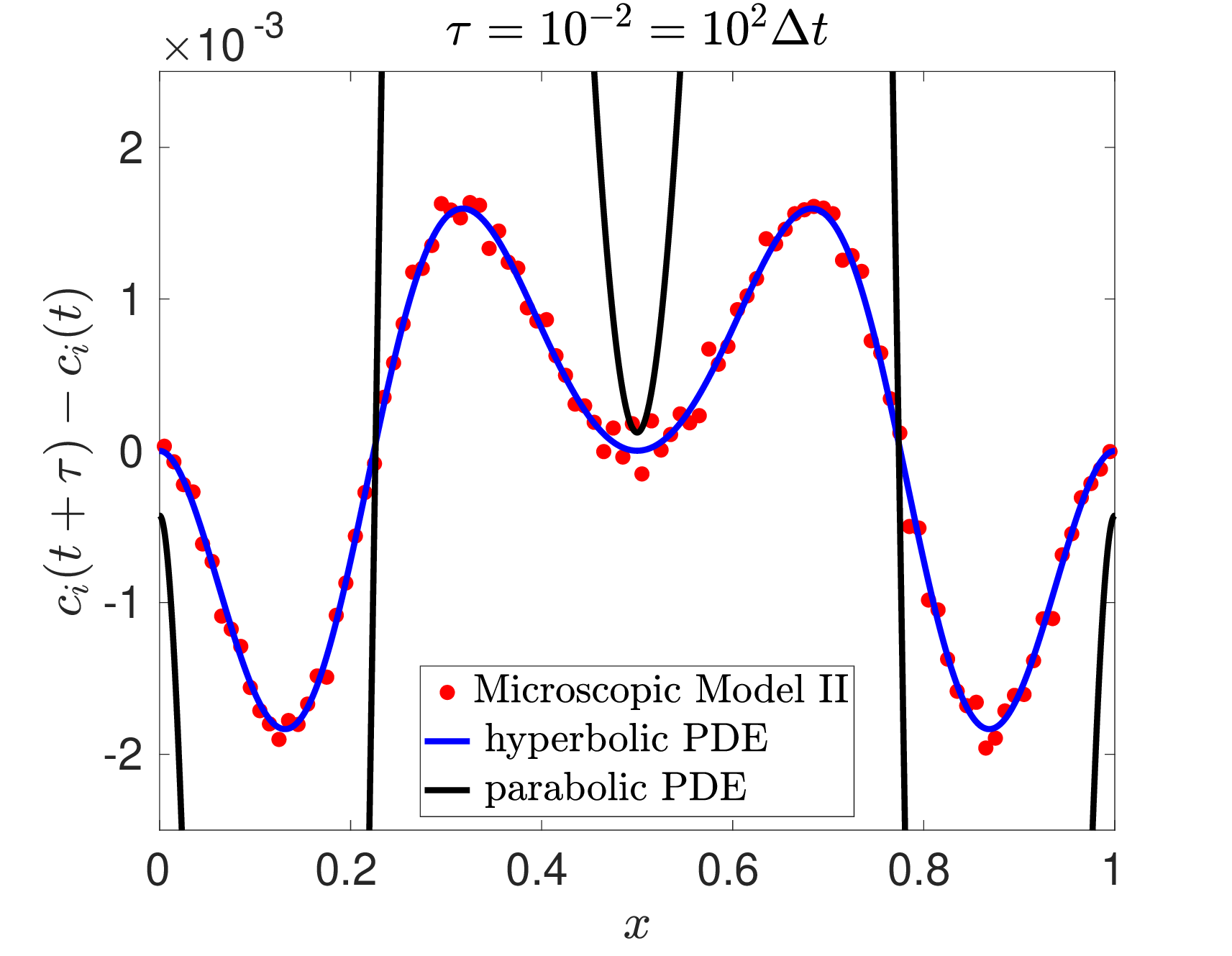,height=5.1cm}
\hskip -2.8mm
\epsfig{file=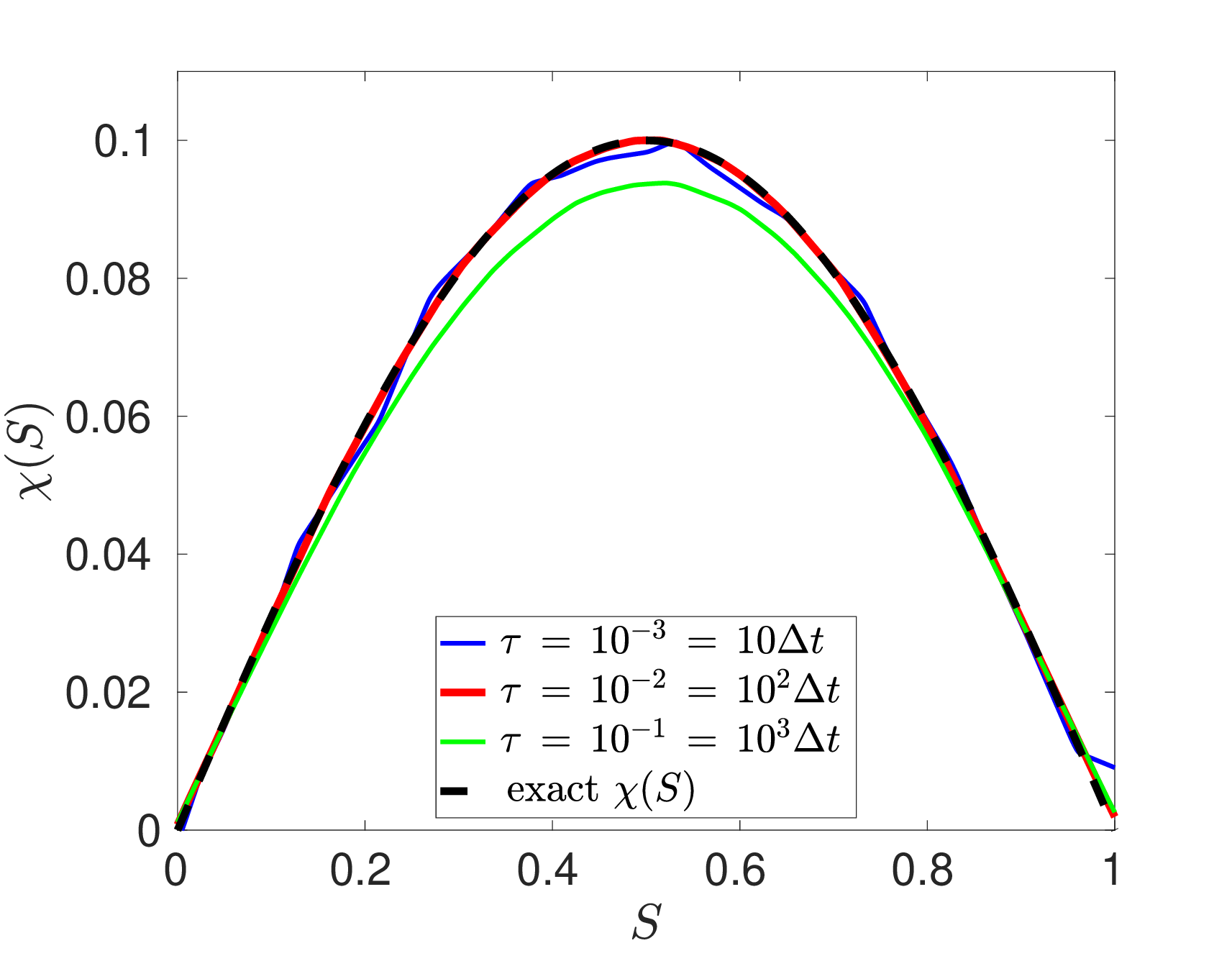,height=5.1cm}
\vskip -5.3cm
(a) \hskip 59mm (b)
\vskip 4.7cm
\caption{{\it {\rm (a)} The change in the concentration $c_i(t+\tau)-c_i(t)$ over the time interval of length~$\tau$, starting from the uniform concentration profile $c_i(t) \equiv 1$ for $i=1,2,\dots,n$, calculated by Microscopic~Model {\rm II} $($red circles$)$. We use chemotactic sensitivity~$(\ref{givenchiS})$, signal profile~$(\ref{signalS})$, $N=10^9$ cells in the microscopic model, time interval $\tau = 10^{-2} = 10^2 \Delta t$, $\beta=0.4$, $\lambda_0 = 8$ and $n=100$. \hfill\break The solutions of the parabolic chemotaxis equation~$(\ref{1Dchemeqn})$ and the hyperbolic chemotaxis equation~$(\ref{csecorder})$ are plotted as the black and blue line, respectively.
 \hfill\break {\rm (b)} The chemotactic sensitivity estimated by the feedforward neural network for different values of the time interval~$\tau$. The exact chemotactic sensitivity~$(\ref{givenchiS})$ is plotted as the black dashed line.
}}
\label{fig3}
\end{figure}

In Figure~\ref{fig3}(a), we present illustrative results calculated by Microscopic Model~II. We consider $N=10^9$ cells, $\beta=0.4$, $\lambda_0 = 8$, $\Delta t = 10^{-4}$, $L=1$ and the signal profile given by~(\ref{signalS}) together with the chemotactic sensitivity $\chi(S)$ given by~(\ref{givenchiS}). In particular, equation~(\ref{Dform}) implies that $D=10^{-2}$, which means that we use the same diffusion constant $D$ and chemotactic signal $S(x)$ as we used in our previous simulation of Microscopic Model~I in Figure~\ref{fig2}(a). We consider the same initial condition for positions, uniform distribution $c_i(t) \equiv 1$, and we initialize a half of the cells with the positive velocity, $+\beta$, and a half of them with the negative velocity, $-\beta$, so that we have $z_i(t) \equiv 0$ as the initial condition of equations~(\ref{hyperchemodiscrci})--(\ref{hyperchemodiscrzi}) at time $t$. We calculate the evolution over the time interval $[t,t+\tau]$, where $\tau = 10^{-2} = 10^2 \Delta t$. We note that this value of $\tau$ is 10-times longer than the value of $\tau$ used in Figure~\ref{fig2}(a). In particular, the solution of the parabolic chemotaxis equation~$(\ref{1Dchemeqn})$ varies more in Figure~\ref{fig3}(a) than in Figure~\ref{fig2}(a). However, it is not the  parabolic chemotaxis equation~(\ref{1Dchemeqn}) but the hyperbolic chemotaxis equation~(\ref{csecorder}), which approximates well Microscopic Model~II. This is illustrated in Figure~\ref{fig3}(a), where some parts of the solution of the parabolic chemotaxis equation~(\ref{1Dchemeqn}) are outside of the plotted range of values.

The solutions of macroscopic PDEs~(\ref{1Dchemeqn}) and~(\ref{csecorder}) converge as $t \to \infty$ to the same steady state solution~(\ref{ststsol}). In particular, if $\tau$ was sufficently large (approximately of the order $\tau = 10 = 10^5 \Delta t$ for our parameter values), then the black and blue solid lines in Figure~\ref{fig3}(a) would be close to each other and close to the steady state solution~(\ref{ststsol2}). However, our neural network estimation uses relatively small values of $\tau$ and Figure~\ref{fig3}(a) highlights a large difference between the transient solutions of macroscopic PDEs~(\ref{1Dchemeqn}) and~(\ref{csecorder}) for $\tau = 10^{-2} = 10^2 \Delta t$. Therefore, we will modify the loss function~(\ref{lossfunction}) to include the central difference approximation of the additional term describing the second time derivative in equation~(\ref{csecorder}). We use
\begin{equation}
{\mathcal L}_2(S)
\,=\,
\frac{1}{n}
\sum_{i=1}^n
\left(
\frac{c_i(t+2\tau) - 2 c_i(t+\tau) + c_i(t)}{2 \lambda_0 \tau^2}
\,+\,
\frac{c_i(t+2\tau) - c_i(t)}{2\tau}
\,-\,
{\mathcal A}_{i} (c,S)
\!\right)^{\!2} \, ,
\label{lossfunction2}
\end{equation}
where the spatial derivative term ${\mathcal A}_{i} (c,S)$ is evaluated using the calculated data $c_i(t+\tau)$ at time $t+\tau$. To estimate the chemotactic sensitivity $\chi(S)$ from relatively short simulations of Microscopic Model~II, we use the feedforward neural network architecture schematically shown in Figure~\ref{fig1}(a) with $n_1=n_2=50$. Our data are calculated as in Figure~\ref{fig3}(a), where we consider that cells are uniformly distributed at time $t$, i.e. $c_i(t) \equiv 1$. We calculate their time evolution by Microscopic Model~II over the time interval $[t,t+2\tau]$ to estimate $c_i(t+\tau)$ and $c_i(t+2\tau)$, where $\tau>0$. We use 
$
\tau \in \big\{ 10 \Delta t, \, 10^2 \Delta t, \, 10^3 \Delta t \big\}
$ 
and train 100 neural networks (using $10^3$ epochs) for each triplet $c_i(t)$, $c_i(t+\tau)$and $c_i(t+2\tau)$. The results are presented in Figure~\ref{fig3}(b), where we average the calculated chemotactic sensitivity $\chi(S)$ over 100 realizations of the training process for each value of $\tau$.

\section{Microscopic Model~III (with internal dynamics)}

\label{sec5}

When bacteria {\it E. coli} move in a favourable direction (that is, in the direction of increasing chemoatractant $S$) the run times are increased and the turning frequency of the velocity jump process decreases. This has been modelled in Section~\ref{sec4} by the turning frequency in the form~(\ref{turnfreq}), which assumes that a cell can directly estimate the gradient of the chemoattractant concentration at its current position. 

In more detailed individual-based models of chemotaxis, {\it E. coli} does not directly estimate the gradient, but only the absolute value of the chemoattractant at its current position, $S(X_i(t))$. {\it E. coli} detects the attractant concentration by receptors on its membrane and the information about the receptor occupancy is then processed by intracellular signalling molecules forming the signal transduction network~\cite{Barkai:1997:RSB,Spiro:1997:MEA}. Therefore, each bacterium is described not only by its position and velocity, but also by additional, internal, variables. Simplified models of this process were studied in~\cite{Erban:2004:ICB,Erban:2005:STS} which showed that models with one or two internal variables can be used to describe the excitation-adaptation properties of the intracellular signal transduction network. We will study such a one-dimensional model in this paper as Microscopic Model~III.

We model a system of $N$ unicellular organisms in domain $\Omega = [0,L]$, where the state of the system at time $t$ is described by three $N$-dimensional vectors: the vector of positions~(\ref{statevector}), the vector of velocities~(\ref{statevectorvelocity}) and the vector of internal variables denoted by
\begin{equation}
\mathbf{Y}(t)
=
\big[Y_1(t), Y_2(t), \dots, Y_N(t)\big]
\in {\mathbb R}^{N},
\label{statevectorinternal}
\end{equation}
where $Y_i(t) \in {\mathbb R}$ is the internal variable of the $i$-th cell at time $t$ for $i=1,2,\dots,N$, which evolves according to the ODE
\begin{equation}
\frac{\mbox{d}Y_i}{\mbox{d}t} = \frac{\psi\big(S(X_i(t))\big) - Y_i}{t_a},
\label{ODEinternal}
\end{equation}
where $\psi^\prime = \chi$ is the integral of the chemotactic sensitivity and $t_a>0$ is the adaptation time of the signal transduction network~\cite{Erban:2004:ICB}. We again assume that each cell moves along the $x$-axis at a constant speed $\beta>0$, i.e. its velocity $V_i(t)$ only takes one of two possible values, $V_i(t) = \pm \beta$. Each cell reverses its direction at random instants of time according to the Poisson process with turning frequency
\begin{equation}
\mbox{ReLU} \! \left( \lambda_0 \, + \,\frac{\lambda_0 \, (1+2 t_a) \, \big( Y_i - \psi(S) \big)}{t_a \, \beta^2}
\right),
\label{turnfreqinternal}
\end{equation}
where $\lambda_0 > 0$, the function $\mbox{ReLU}$ is defined by~(\ref{defReLU}) and signal $S$ is evaluated at the current position of the cell, $X_i(t)$. Since the equation~(\ref{ODEinternal}) is solved along the trajectory, the biological meaning of internal variables is that they form a primitive chemical ``memory" enabling a comparison of the cell's current environment with the environment it visited a while ago. The turning frequency~(\ref{turnfreqinternal}) is chosen as the constant $\lambda_0$ plus the second term which depends on the internal variable $Y_i$. This is similar to the turning frequency~(\ref{turnfreq}). If the concentration of chemoattractant~$S$ is increasing along the cellular trajectory, then the bacterium is moving in a good direction and it is less likely to change it. On the other hand, if the concentration of~$S$ is decreasing along its trajectory, then the bacterium is more likely to turn. 

To simulate Microscopic Model~III, we update the system over time steps of length~$\Delta t$. Multiplying the turning frequency~(\ref{turnfreqinternal}) by $\Delta t$ and assuming that $\Delta t$ is sufficiently small, the probability of turning during one time step is given by
\begin{equation}
p_i(t)
=
\mbox{ReLU} \! \left( \lambda_0 \, + \,\frac{\lambda_0 \, (1+2 t_a) \, \big( Y_i - \psi\big(S(X_i(t))\big) \big)}{t_a \, \beta^2} \right) \Delta t \, .
\label{probturning2}
\end{equation}
In our simulations, we generate $N$ random numbers $r_i$, $i=1,2,\dots,N$, uniformly distributed in interval $[0,1]$, and the $i$-th cell changes its direction of movement at time $t$ if $r_i < p_i(t)$. Then we update the position $X_i(t)$ of each cell by~(\ref{positionupdate}). To calculate $Y_i(t+\Delta t)$, we need to solve the ODE~(\ref{ODEinternal}) over the time interval~$[t,t+\Delta t]$. Since $V_i(t)$ is assumed to be constant during the time interval $[t,t+\Delta t$], equation~(\ref{ODEinternal}) reduces to solving
\begin{equation}
\frac{\mbox{d}Y_i}{\mbox{d}\sigma} = \frac{\psi\big(S(X_i(t)+\sigma V_i(t))\big) - Y_i}{t_a},
\qquad \mbox{for} \; \sigma \in [0,\Delta t],
\label{ODEinternalsigma}
\end{equation}
given the initial condition $Y_i(t)$ for $\sigma=0.$ 

Equation~(\ref{ODEinternalsigma}) can be solved by a numerical method for solving ODEs, which is often the case when using complex ODE models of signal transduction networks~\cite{Barkai:1997:RSB,Spiro:1997:MEA}. In our case, we only have one single ODE~(\ref{ODEinternalsigma}), so we can also proceed with an analytic approach, which will highlight the hidden dependence of Microscopic Model~III on the signal derivative. To see this, we observe that $\sigma  \in [0,\Delta t]$ is small and we use the Taylor expansion to approximate
$$
\psi\big(S(X_i(t)+\sigma V_i(t))\big)
\approx
\psi\big(S(X_i(t))\big)
+
\chi\big(S(X_i(t))\big)
\, 
\frac{\partial S}{\partial x} (X_i(t)) 
\, \sigma \, V_i(t) \,,
$$
where $\psi^\prime = \chi$ and we neglected terms of the order ${\mathcal O}(\sigma^2).$ Substituting into~(\ref{ODEinternalsigma}) and solving the resulting linear ODE, we obtain
\begin{eqnarray}
Y_i(t+\Delta t) 
&=& 
Y_i(t) \exp \left( - \frac{\Delta t}{t_a} \right)
+
\psi\big(S(X_i(t))\big)
\left(
1
-
\exp \left( - \frac{\Delta t}{t_a} \right)
\right)
\nonumber
\\
&+&
\chi\big(S(X_i(t))\big)
\, 
\frac{\partial S}{\partial x} (X_i(t)) 
\, V_i(t) 
\left(
\Delta t
-
t_a 
+
t_a
\exp \left( - \frac{\Delta t}{t_a} \right)
\right),
\label{ytpdtupdate}
\end{eqnarray}
which we use in our implementation of Microscopic Model~III to calculate the update of the internal variable over one time step $[t,t+\Delta t]$. The formula~(\ref{ytpdtupdate}) not only gives us a possible way how we can calculate $Y_i(t+\Delta t)$ at time $t+\Delta t$ given the values $X_i(t),$ $V_i(t)$ and $Y_i(t)$ at time $t$, but it also illustrates that cells effectively extract some information on the chemotactic sensitivity $\chi = \psi^\prime$ and the signal derivative even if the formulation of Microscopic Model~III assumes that cells can only measure the absolute values of the signal $S$ in their environment. 

\begin{figure}
\rule{0pt}{1pt}
\vskip 1mm
\epsfig{file=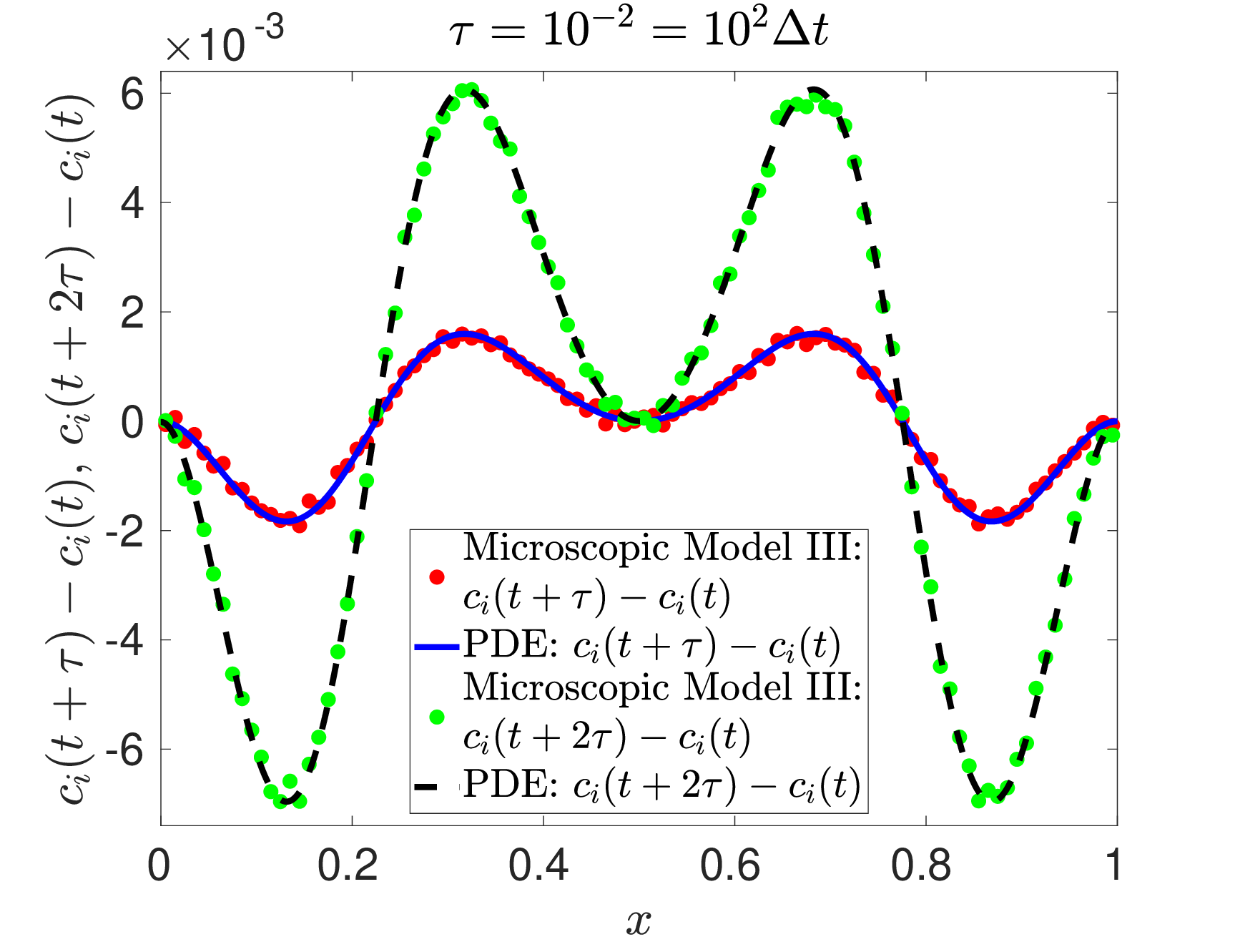,height=5.1cm}
\hskip -2.8mm
\epsfig{file=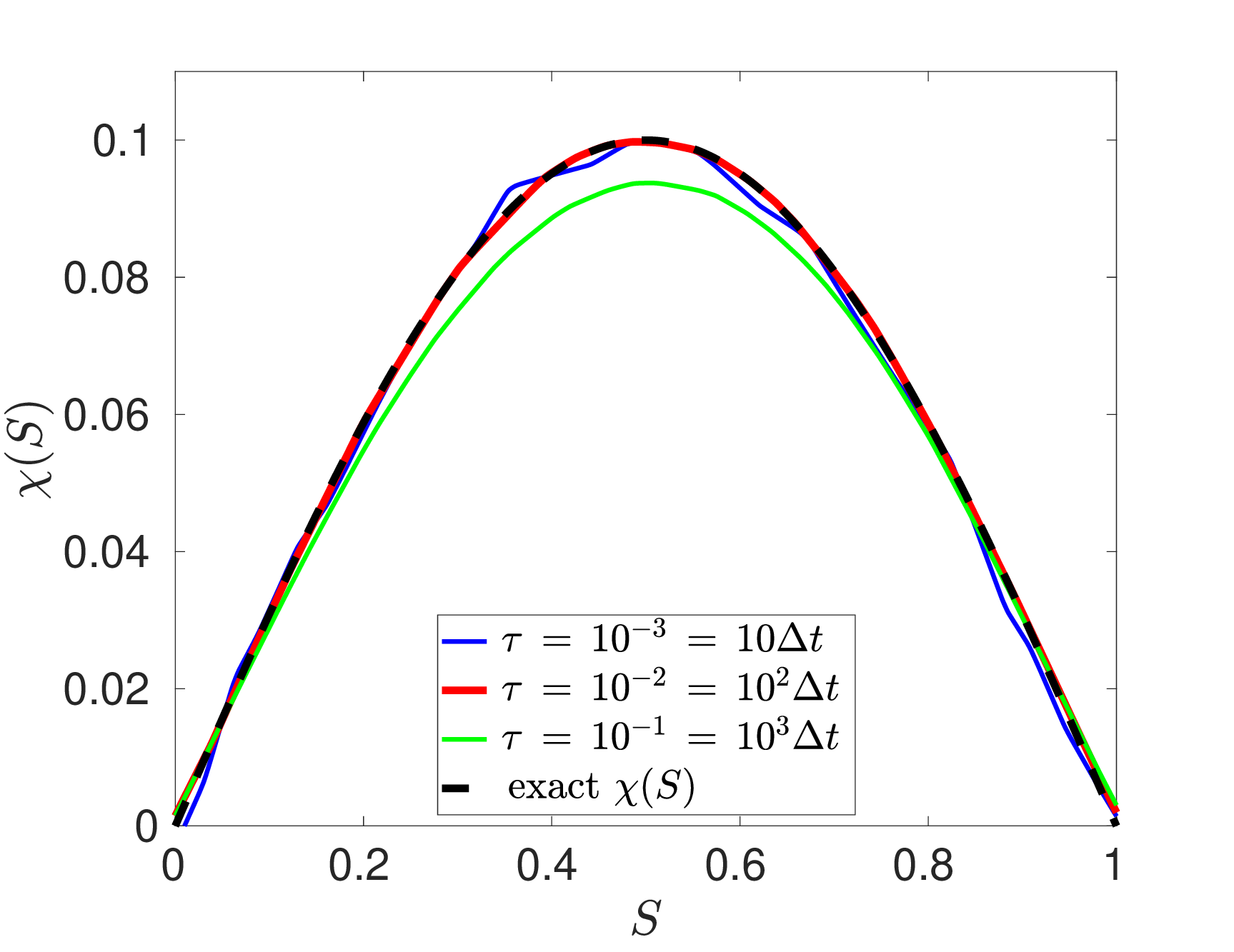,height=5.1cm}
\vskip -5.3cm
(a) \hskip 59mm (b)
\vskip 4.7cm
\caption{{\it {\rm (a)} The changes in the concentration $c_i(t+\tau)-c_i(t)$ and $c_i(t+2\tau)-c_i(t)$ over the time intervals of length $\tau$ and $2\tau$, respectively, starting from the uniform concentration profile $c_i(t) \equiv 1$ for $i=1,2,\dots,n$, calculated by Microscopic Model~{\rm III} $($circles\,$)$. We use chemotactic sensitivity~$(\ref{givenchiS})$, signal profile~$(\ref{signalS})$, $N=10^9$ cells in the microscopic model, time interval $\tau = 10^{-2} = 10^2 \Delta t$, $\beta=0.4$, $\lambda_0 = 8$, $t_a = 10^{-4}$ and $n=100$. The solutions of the hyperbolic chemotaxis equation~$(\ref{csecorder})$ are plotted as the solid blue and dashed black lines, respectively.
 \hfill\break {\rm (b)} The chemotactic sensitivity estimated by the feedforward neural network for different values of the time interval $\tau$. The exact chemotactic sensitivity~$(\ref{givenchiS})$ is plotted as the black dashed line.
}}
\label{fig4}
\end{figure}

Using~\cite[equation (6.34)]{Erban:2004:ICB}, we can also derive the hyperbolic chemotaxis equation in the form~(\ref{csecorder}), which provides good approximation of the macroscopic behaviour of Microscopic Model~III for some parameter regimes. This is illustrated in Figure~\ref{fig4}(a), where we present illustrative results calculated by Microscopic Model~III. We consider $N=10^9$ cells, $\beta=0.4$, $\lambda_0 = 8$, $\Delta t = 10^{-4}$, $t_a = 10^{-4}$, $L=1$ and the signal profile given by~(\ref{signalS}) together with the chemotactic sensitivity $\chi(S)$ given by~(\ref{givenchiS}). In particular, equation~(\ref{Dform}) implies that $D=10^{-2}$, which means that we use the same diffusion constant $D$, turning frequency $\lambda_0$ and chemotactic signal $S(x)$ as we used in our previous simulation of Microscopic Model~II in Figure~\ref{fig3}(a). We also consider the same initial condition for positions and velocity, namely uniform distribution $c_i(t) \equiv 1$, and we initialize a half of the cells with the positive velocity, $+\beta$, and a half of them with the negative velocity, $-\beta$, so that we have $z_i(t) \equiv 0$ as the initial condition of equations~(\ref{hyperchemodiscrci})--(\ref{hyperchemodiscrzi}) at time $t$. To initialize internal dynamics variables, we put $Y_i(0)=\psi\big(S(X_i(0))\big)$. We calculate the evolution over the time interval $[t,t+2\tau]$, where $\tau = 10^{-2} = 10^2 \Delta t$ and we plot both $c_i(t+\tau)-c_i(t)$ and $c_i(t+2\tau)-c_i(t)$ in Figure~\ref{fig4}(a), which are the training data needed for our neural network estimation of the chemotactic sensitivity when using the loss function ${\mathcal L}_2(S)$ given by~(\ref{lossfunction2}).

In Figure~\ref{fig4}(b), we present results of the estimation of the chemotactic sensitivity $\chi(S)$ from relatively short simulations of Microscopic Model~III, using the feedforward neural network architecture schematically shown in Figure~\ref{fig1}(a) with $n_1=n_2=50$. Our data are calculated as in Figure~\ref{fig4}(a), where we consider that cells are uniformly distributed at time $t$, i.e. $c_i(t) \equiv 1$. We calculate their time evolution by Microscopic Model~III over the time interval $[t,t+2\tau]$ to estimate $c_i(t+\tau)$ and $c_i(t+2\tau)$, where $\tau>0$, which we use in the loss function~(\ref{lossfunction2}). Using 
$
\tau \in \big\{ 10 \Delta t, \, 10^2 \Delta t, \, 10^3 \Delta t \big\}\,,
$ 
we train 100 neural networks (using $10^3$ epochs) for each triplet $c_i(t)$, $c_i(t+\tau)$ and $c_i(t+2\tau)$. The results shown in Figure~\ref{fig4}(b) present the average calculated chemotactic sensitivity $\chi(S)$ over 100 realizations of the training process for each value of $\tau$.

\section{Discussion}

\label{sec6}

Movement of individual biological agents often depends on external signal fields (for example, cells are attracted to nutrients and repelled by toxins), and population-level mathematical models of their behaviour have traditionally been formulated in terms of PDEs, such as the parabolic chemotaxis equation~(\ref{chemotaxisequation}). At the individual level, stochastic models have been used, including Brownian dynamics and velocity-jump processes, and an important mathematical question has been to connect these two levels of description~\cite{Othmer:1997:ABC}. In this paper, we have investigated three individual-based models, denoted Microscopic Models~I, II and~III. In each case, the cellular population (in the limit of many individuals, $N \to \infty$) can be described by a classical chemotaxis equation of the form~(\ref{chemotaxisequation}). However, it is only for the simplest Microscopic Model~I that the limit $N \to \infty$ can be established analytically for all parameter regimes, and the individual-based description, formulated as Brownian dynamics with finite $N$ and a finite time step $\Delta t$, converges to the PDE description in the limit $N \to \infty$ and $\Delta t \to 0$.

Microscopic Models~II and~III are formulated as velocity-jump processes in which the turning frequency depends on the extracellular signal $S$, either directly, as in~(\ref{turnfreq}), or indirectly through internal dynamics, as in (\ref{turnfreqinternal}). Then the classical chemotaxis equation can only be analytically derived under certain additional assumptions, for example, when the signal gradients, are sufficiently small~\cite{Erban:2004:ICB,Xue:2015:MEB}. Microscopic Model~III has included a simplified description of internal dynamics, using only one internal variable $Y_i(t)$ describing the adaptation dynamics. To model the excitation-adaptation dynamics of the intracellular signal transduction networks~\cite{Othmer:1998:OCS,Erban:2005:STS}, we can use a model with two variables $Y_i(t)$ and $Z_i(t)$
\begin{equation}
\frac{\mbox{d}Z_i}{\mbox{d}t} = \frac{g(C(t)) - Z_i - Y_i}{t_e},
\qquad
\frac{\mbox{d}Y_i}{\mbox{d}t} = \frac{g(C(t)) - Y_i}{t_a},
\label{cartooninternalC}
\end{equation}
where $C(t) = S({\mathbf X}_i(t),t)$ is the signal detected by a cell at its current position ${\mathbf X}_i(t)$, function $g: [0,\infty) \to [0,\infty)$ incorporates sensitivity of the signal transduction network to different levels of the detected signal and constants $t_e$ and $t_a$ are the excitation time and the adaptation time, respectively. If $C(t)$ is a constant, then the steady state value of $Z_i$ is independent of this constant, i.e. the values of $Z_i$ perfectly adapt to any constant stimulus, and the velocity-jump description of chemotaxis can be formulated by assuming that the turning frequency depends on $Z_i$. For example, if the turning frequency is given by $\mbox{ReLU} (\lambda_0 - b_0 Z_i)$ for constants $\lambda_0>0$ and $b_0>0$, then model~(\ref{cartooninternalC}) leads to the hyperbolic chemotaxis equation of the form~(\ref{csecorder}), provided that the signal gradients are sufficiently small, with the chemotactic sensitivity given by~\cite{Erban:2004:ICB,Erban:2005:STS}
$$
\chi(S)
=
\frac{b_0 \, \beta^2 \; \! t_a \, g^\prime(S)}
{\lambda_0 (1 + 2 \lambda_0 t_a)(1 + 2 \lambda_0 t_e)}\,.
$$
Since the adaptation time $t_a$ is much larger than the excitation time $t_e$, one can also work with the simplifed model, where $t_e = 0$, which has been our case of Microscopic Model~III. We have studied one-dimensional models, but Microscopic Models~I, II and~III can be formulated in two and three spatial dimensions, and it is possible to establish analytically links with the chemotaxis equation~(\ref{chemotaxisequation}), but the derivation is more technical~\cite{Hillen:2001:TEC,Erban:2005:STS} and the simplifying assumptions of shallow signal gradients
must again be imposed for Microscopic Models~II and~III. This analysis can be further extended to velocity-jump processes which assume that the change in the direction of motion is not instantaneous, including models of the tumbling phase of {\it E. coli} motion~\cite{Erban:2004:ICB}, and other delays caused by turning of individuals~\cite{Taylor:2015:MMT}.

More complex chemotaxis models may include dozens of internal variables representing the concentrations of intracellular signalling molecules involved in signal transduction networks~\cite{Barkai:1997:RSB,Spiro:1997:MEA}, or written as PDEs describing spatio-temporal concentrations of signalling molecules~\cite{Erban:2007:TEA,Giniunaite:2020:MCC}. Some models also incorporate interactions between individuals, which can be direct -- via volume exclusion, whereby only a finite number of cells can occupy a given volume -- or indirect, by modifying the extracellular environment through the release or consumption of signalling molecules. In this case, the time evolution of extracellular signals must also be modelled. A number of individual-based models of interacting cells have been developed in the literature~\cite{Dallon:1997:DCM,Franz:2012:HMI,Franz:2014:TWH,Yasuda:2021:EID}. The key aspect of these hybrid models is the simulation of cells as individual particles while describing the extracellular environment by PDEs. The same situation also arises when using Brownian dynamics of charged particles, where models of individual particles need to be coupled with the description of electrostatic potential solving the Poisson equation~\cite{Hockney:1988:CSU,Zhang:2025:BDP}. Some models of cell migration also include domain growth, which introduces additional terms into population-level PDEs~\cite{Baker:2010:MMD,Macdonald:2016:CMC,Krause:2023:CDE}. Considering complex chemotaxis models, computational frameworks for extracting macroscopic behaviour from individual-based descriptions are often necessary in some parameter regimes, where macroscopic equations cannot be analytically derived~\cite{Erban:2006:EFC,Lee:2023:LBG,Psarellis:2024:DDC}.

In this paper, we have used feedforward neural networks to estimate macroscopic chemotactic sensitivity $\chi(S)$ using data calculated by relatively short simulations of Microscopic Models~I, II and~III. While the parabolic chemotaxis equation~(\ref{1Dchemeqn}) provides a good description of macroscopic dynamics corresponding to Microscopic Model~I on such short time scales, this is not the case when more complex Microscopic Models~II and~III are used. The convergence of Microscopic Model~I to the parabolic chemotaxis equation~(\ref{1Dchemeqn}) is illustrated in Figure~\ref{fig2}(a), while we can see large quantitative differences in transient dynamics between the parabolic chemotaxis equation~(\ref{1Dchemeqn}) and Microscopic Model~II on short time scales in Figure~\ref{fig3}(a). Consequently, the neural network estimation with loss function~${\mathcal L}(S)$ given by~(\ref{lossfunction}) works well in the case of Microscopic Model~I, but we need to take into account different transient dynamics of velocity-jump processes and use loss function~${\mathcal L}_2(S)$ given by(\ref{lossfunction2}) when working with short-time simulations of Microscopic Models~II and~III. 

Figures~\ref{fig3}(b) and~\ref{fig4}(b) illustrate that the loss function~(\ref{lossfunction2}) works well for short simulation time windows, with the best performance obtained for $\tau = 10^{-2} = 10^2 \Delta t$. If we consider shorter time windows, then the results are influenced by noise stemming from the estimation of densities in simulations of a finite number of individuals, $N$, as we can see for $\tau = 10^{-3} = 10 \Delta t$ in Figures~\ref{fig3}(b) and~\ref{fig4}(b). On the other hand, if $\tau$ is chosen larger than optimal, then the time derivatives in the loss function~(\ref{lossfunction2}) will be approximated with a bias leading to the error seen for $\tau = 10^{-1} = 10^3 \Delta t$ in Figures~\ref{fig3}(b) and~\ref{fig4}(b). The same conclusion can be made for Microscopic Model~I in Figure~\ref{fig2}(b) with the best performing value of the time window being~$\tau = 10^{-3} = 10 \Delta t$. This is shorter than for velocity-jump processes because the transient dynamics of the parabolic chemotaxis equation~(\ref{1Dchemeqn}) is faster than the hyperbolic chemotaxis equation~(\ref{csecorder}) for our parameters and initial conditions. The solutions of macroscopic PDEs~(\ref{1Dchemeqn}) and~(\ref{csecorder}) converge as $t \to \infty$ to the same steady state solution~(\ref{ststsol}). In particular, if a model is sufficiently simple that long-time equilibrium simulations of an individual-based model are computationally feasible, then they could also be used for estimating the chemotactic sensitivity $\chi(S)$.

\vskip 4mm

\noindent
{\bf Funding}: This work was supported by the Engineering and Physical Sciences Research Council, grant number EP/V047469/1.

\end{document}